\documentclass[12pt]{article}
\usepackage{graphicx}

\def\napoli{Institute of Physics, Czech Academy of Sciences, 18221 Prague, Czech Republic}

\def\Title#1{\begin{center} {\Large #1 } \end{center}}
\def\Author#1{\begin{center}{ \sc #1} \end{center}}
\def\Address#1{\begin{center}{ \it #1} \end{center}}

\newenvironment{Abstract}{\begin{quotation}  }{\end{quotation}}
\newenvironment{Presented}{\begin{quotation} \begin{center} 
             PRESENTED AT\end{center}\bigskip 
      \begin{center}\begin{large}}{\end{large}\end{center} \end{quotation}}

\def\beq{\begin{equation}}
\def\eeq#1{\label{#1}\end{equation}}
\def\eeqn{\end{equation}}

\def\beqa{\begin{eqnarray}}
\def\eeqa#1{\label{#1}\end{eqnarray}}
\def\eeqan{\end{eqnarray}}

\let\bar=\overbar

\def\Dslash{\not{\hbox{\kern-4pt $D$}}}
\def\dslash{\not{\hbox{\kern-2pt $\del$}}}

\def\msb{{\bar{\ssstyle M \kern -1pt S}}}

\begin{document}
\begin{titlepage}
%\pubblock

\vfill
\Title{Soft Particle Production in Cosmic Ray Showers}
\vfill
\Author{Jan Ebr}
\Address{\napoli}
\vfill
\begin{Abstract}
Indications of a discrepancy between simulations and data on the number of muons in cosmic ray showers exist over a large span of energies. We focus on the excess of multi-muon bundles observed by the DELPHI detector at LEP and on the excess in the muon number in general reported by the Pierre Auger Observatory. Even though the primary CR energies relevant for these experiments differ by orders of magnitude, we can find a single mechanism which can simultaneously increase predicted muon counts for both, while not violating constraints from accelerators or from the longitudinal shower development as observed by the Pierre Auger Observatory. We present a brief motivation and describe a practical implementation of such a model, based on the addition of soft particles to interactions above a chosen energy threshold. Results of an extensive set of simulations show the behavior of this model in various parts of a simplified parameter space.
\end{Abstract}
\vfill
\begin{Presented}
Presented at EDS Blois 2017, Prague, \\ Czech Republic, June 26-30, 2017
\end{Presented}
\vfill
\end{titlepage}
\def\thefootnote{\fnsymbol{footnote}}
\setcounter{footnote}{0}

\section{Motivation}

Cosmic ray showers allow us to not only study astrophysical phenomena, but also to test the interaction models used in particle physics. It is well known that the ultra-high energy cosmic rays with primary energies above $10^{17}$~eV provide the unique opportunity to study hadronic interactions at energies ($\sqrt{s}$) larger than those achieved by the LHC, however also intermediate-energy cosmic rays at energies around $10^{16}$~eV can be of particle physics interest as fixed-target experiments do not even remotely approach such energies and the kinematic coverage of detectors in colliders is incomplete -- in fact it is such that the forward-backward regions most significant for the development of the cosmic rays showers are the least well studied at the colliders, despite numerous dedicated efforts at the LHC.

The importance of the study of cosmic ray showers for hadronic physics is made clear by the existing discrepancies between models and data, typically regarding the amount or energy spectrum of muons observed at or under the ground level. At the ultra-high energy end, the most prominent are the results of the Pierre Auger Observatory \cite{auger}. At intermediate energies, multiple indications of such discrepancies exist, however a particularly intriguing one is that of the DELPHI detector at the LEP collider \cite{delphi}, which was endowed with a significant rock overburden allowing the separation of very energetic muons (with vertical cutoff of 52 GeV). The reported discrepancy there is in the number of observed muon bundles of large multiplicities.

The discrepancy reported by Auger comes from comparison with current hadronic interaction models, but the DELPHI results are a decade old. We have thus first checked what effect would the application of recent models in the analysis of DELPHI data have using a simplified model of the detector. The results are such that using QGSJET-II-04 \cite{qgsjet} reduces the gap between the predicted number of multi-muon bundles and the observed one by 35 to 15 \% depending on the multiplicity region in question while using EPOS-LHC \cite{epos} actually makes the discrepancy even slightly larger and thus the effect is by no means explained by the current generation of interaction models.

While the Auger muon excess is observed at energies around $10^{19}$~eV, the dominant contribution to the muon bundles observed at DELPHI comes from showers with energies in the $10^{15}$ to $10^{17}$~eV range. It is not clear that the two are caused by the same physical effect, but we explore the implication of the assumption that they are. With such an assumption, the only mechanism that we were able to find that has the desired effect on the simulations in both cases is the addition of particles of very small momenta (in hundreds of MeV) in the center-of-mass frame of the collision; we call those ``soft particles'' for brevity and the details of the model and its results can be found in \cite{spam}. Here we present the main highlights of the work.

%%%%%%%%%%%%%%%%%%%%%%%%%%%%%%%%%%%%%%%%%%%%%%%%%%%%%%%%%%%%%%%%%%%%%%%%%
%%
%% use this format to include an .eps figure into your paper
%%
\begin{figure}[htb]
\begin{center}
\includegraphics[height=3.7in]{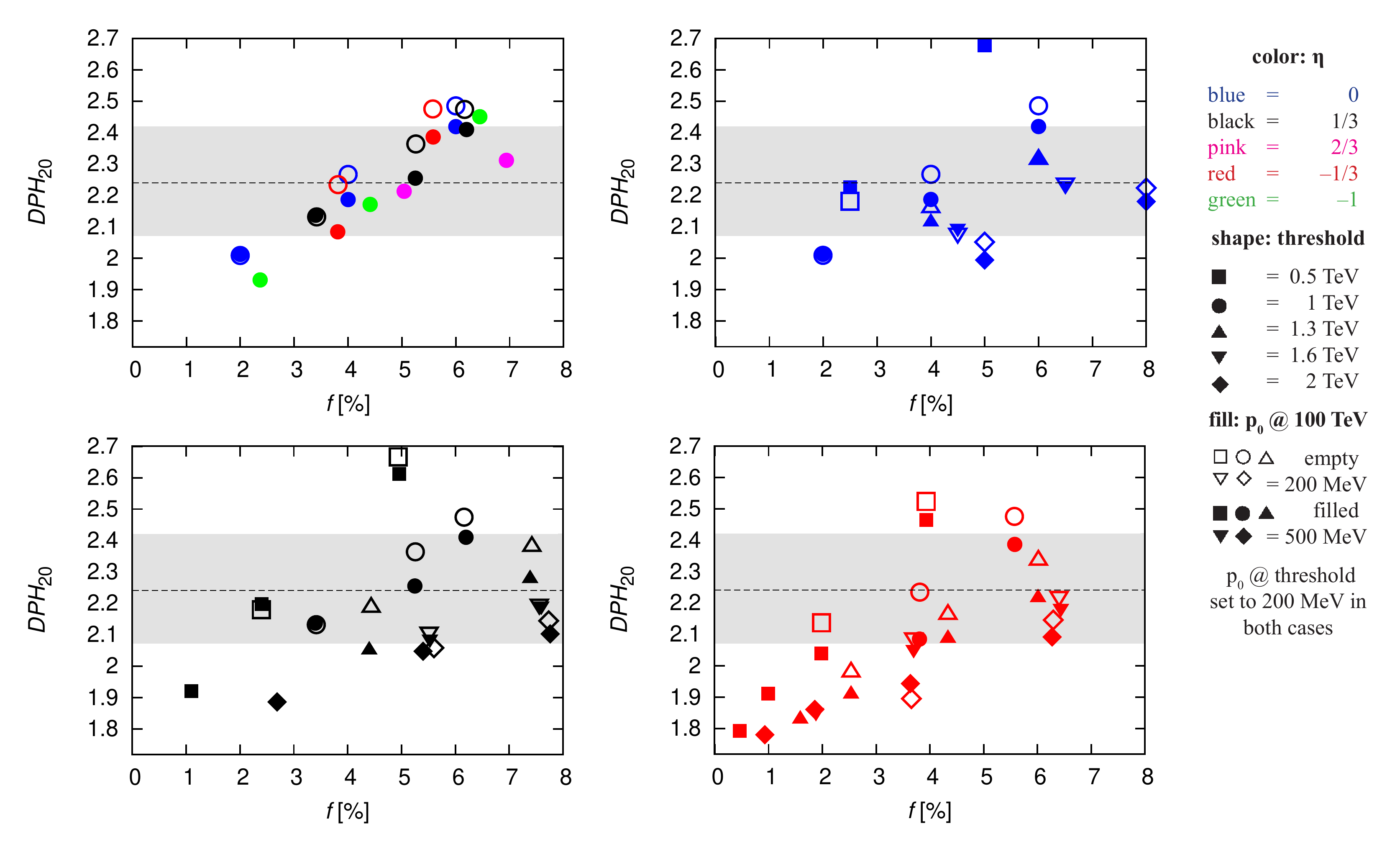}
\caption{The relative number of muon bundles with multiplicity higher than 20 with respect to the QGJSET-01 proton prediction ($DPH_{20}$) vs. the effective average energy fraction converted to soft particles for various choices of model parameters (see text). The horizontal dashed line and the gray band represent the measured DELPHI value and its 1-sigma uncertainty.}
\label{fig:delphi}
\end{center}
\end{figure}
%%%%%%%%%%%%%%%%%%%%%%%%%%%%%%%%%%%%%%%%%%%%%%%%%%%%%%%%%%%%%%%%%%%%%%%%%%%

\section{Model}

\begin{figure}[htb]
\begin{center}
\includegraphics[height=3.7in]{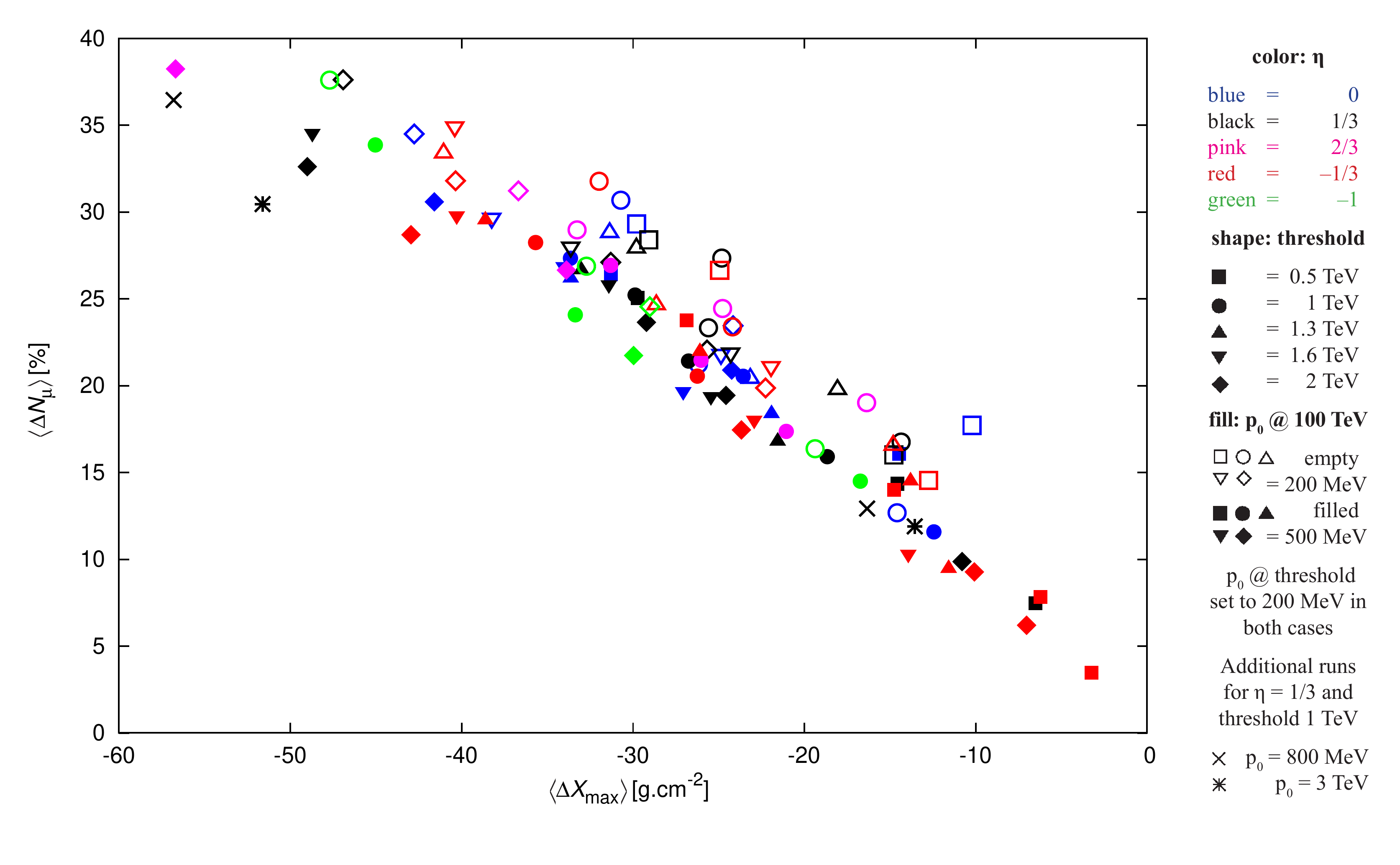}
\caption{The shift in the depth of the shower maximum ($\Delta X_\mathrm{max}$) vs. the change in the number of muons at ground 1000 meters from the shower core ($\Delta N_\mu$) for various choices of model parameters (see text) for proton-induced showers with primary energy of $3.2\times10^{18}$ eV. The individual simulations also differ in the converted energy fraction $f_\mathrm{eff}$.}
\label{fig:auger}
\end{center}
\end{figure}

The parameter space for adding soft particles to interactions is large, but the results for observable parameters usually depend only weakly on the specifics. In our model, we take an interaction performed in a standard way by the QGSJET-II-04 model and add to it a mixture of light mesons and nucleons, so that the representation of each particle type follows typical collider data and all quantum numbers, including isospin, are conserved. The particles are added only within 0.1 degree from the collision axis to avoid detection constraints at the LHC and only to interactions for which $\sqrt{s}$ exceeds a threshold chosen between 0.5 and 2.0 TeV (indicated by the shape of the symbol int he plots). The momentum distribution of the added particles follows $p \exp(-p/p_0)$ where $p_0$ is either set to 200 MeV (empty symbols in the plots) or to 200 MeV at the energy threshold, logarithmically increasing to 500 MeV at $\sqrt{s}=100$~TeV (filled symbols); two other higher choices of $p_0$ were tested with strongly diminishing effect. The amount of particles added is given as a fraction $f$ of the available energy $\sqrt{s}$. Because a dependence on the centrality of the interaction can be chosen (as a power-law dependence on the sum of wounded nucleons in the target and the projectile, coded by color in the plots), an effective average value $f_\mathrm{eff}$ is used instead. Energy and momentum are conserved by removing an appropriate amount of momentum from the particles produced by QGSJET, but only from those with sufficient momenta and proportionally to the magnitude of these momenta.

\section{Results}

Fig.~\ref{fig:delphi} shows that the observed number of bundles with multiplicity over 20 can be reproduced for many different combinations of model parameters within 2 to 6 per cent of energy on average converted to soft particles. Fig.~\ref{fig:auger} shows that the increase in the muon number at ground is proportional to the shift in the predicted depth of shower maximum $X_\mathrm{max}$. This trend is generally expected, because even for the standard hadronic interaction models, shallower showers tend to produce more muons; however the independence of the correlation on the chosen parameters is striking. Furthemore, the correlation between the predictions for DELPHI muon bundles and for the shift in $X_\mathrm{max}$ was studied, because at $3.2\times10^{18}$ eV a shift in prediction of more than roughly 30 gcm$^{-2}$ would be at odds with other Auger data. It turns out that there are many combinations of parameters for which the observed values for muon bundles at DELPHI are reached while $\Delta X_\mathrm{max}<30 \mathrm{gcm}^{-2}$. This is obviously not a proof of the validity of the model, but an indication of its viability as one of the alternatives to explain the observed muon excesses. Another interesting feature of the model is that for most parameter choices, it predicts the proton showers to be shallower while the iron showers are {\it deeper}, thus bringing those closer to each other. Should the model be valid, the capability to infer the primary composition of cosmic rays from the shower properties would be smaller than currently believed.

\bigskip
This work was supported by the grants of the Ministry of Education of the Czech Republic MSMT LM2015038, LG15014 and EU-MSMT CZ.02.1.01/0.0/0.0/16\_013/\\0001402.

\end{document}